\documentclass[rnote]{aa}
\usepackage{graphicx}
\usepackage{graphics}
\usepackage{amssymb}
\usepackage{txfonts}

\begin{document}

\title{On-sky observations with an achromatic hybrid phase knife\\
 coronagraph in the visible}
\titlerunning{On-sky tests of a phase knife coronagraph}

\author{
  L.~Abe\inst{1}\and
  M.~Beaulieu\inst{1}\and
  F.~Vakili\inst{1}\and
  J.~Gay\inst{2}\and
  J.-P.~Rivet\inst{3}\and
  S.~Dervaux\inst{2}\and
  A.~Domiciano~de~Souza\inst{1,2}
}
\authorrunning{L.~Abe et al.}

\institute{
Laboratoire Universitaire d'Astrophysique de Nice (LUAN), CNRS UMR
6525, Parc Valrose, 06108 Nice Cedex 02, France \and
Laboratoire Gemini, CNRS UMR 6203, Observatoire de la C{\^o}te
d'Azur, B.P. 4229, 06304 Nice cedex 4, France \and
Laboratoire Cassiop{\'e}e, CNRS UMR 6202, Observatoire de la C{\^o}te
d'Azur, B.P. 4229, 06304 Nice cedex 4, France
}

\offprints{L.~Abe, \email{abe@optik.mtk.nao.ac.jp}}

\date{Received 7 March 2006 / Accepted 11 August 2006}

\abstract%
{The four-quadrant phase mask stellar coronagraph, introduced by D.~Rouan et al., is capable of achieving very
high dynamical range imaging and was studied in the context of the direct detection of extra-solar planets. Achromatic four-quadrant phase mask is currently being developed for broadband IR applications.}%
{We report on laboratory and on-sky tests of a prototype coronagraph in the visible. This prototype,
the \emph{achromatic hybrid phase knife coronagraph}, was derived from the
four-quadrant phase mask principle.}%
{The instrumental setup implementing the coronagraph itself was designed to record the pre- and
post-coronagraphic images simultaneously so that an efficient real-time image selection procedure
can be performed. We describe the coronagraph and the associated tools that enable robust and
repeatable observations. We present an algorithm of \emph{image selection} that has been tested
against the real on-sky data of the binary star \object{HD~80\,081} (*~38~Lyn).}%
{Although the observing conditions were poor, the efficiency of the proposed method is proven. From this
experiment, we derive procedures that can apply to future focal instruments associating adaptive optics and
coronagraphy, targeting high dynamic range imaging in astronomy, such as detecting extra-solar planets.}%
{}


\maketitle

        \section{Introduction}
        \label{sec:Intro}

The imagery of the faint structures of astronomical objects at high angular resolution (close to 
the diffraction limit) is a difficult challenge. The current ultimate goal is to directly observe
extra-solar planets, whose existence has been unambiguously confirmed by Mayor \& Queloz (\cite{mayor1995}).
Since then, high dynamic range imaging techniques and subsequent instrumental innovations have triggered
the development of ambitious ground-based and space-borne projects to reach this challenging exo-planet
direct-detection goal. In this context, stellar coronagraphy has emerged with the proposal of numerous
coronagraphic techniques -- e.g. AIC (Gay et al. \cite{gay1997}), phase mask
(Roddier et al. \cite{roddier1997}), four quadrant and phase knife (Rouan et al. \cite{rouan2000};
Abe et al. \cite{abe2001a}), binary-shaped masks (Spergel et al. \cite{spergel2001}; Vanderbei et al.
\cite{vanderbei2004}), pupil apodized Lyot (Aime et al. \cite{aime2002}), two-mirror apodization (Guyon \cite{guyon2003}), band-limited coronagraphs (Kuchner et al. \cite{kuchner2002}) -- all capable, in theory,
of being efficient enough to perform direct detection at the $10^9$ contrast ratio. Nevertheless, further
studies led to the conclusion that the main limitations to achieving this performance do not come from the
high dynamic range imaging concept itself, but rather from all the opto-mechanical environment.

Contrary to the instrument design considerations concerning the rejection of undesired scattered-light,
the aspect of observing strategies and image analysis/interpretation in coronagraphy is not taken
sufficiently into account. Coronagraphic imaging is a highly non-linear process, resulting in a
point-spread-function (PSF) variant by translation. As a consequence, interpreting images requires
a high level of knowledge of its instrumental environment (where many kind of calibration paths are
needed, resulting in a global loss of throughput at the level of the science camera).

The purpose of this paper is twofold. First, we describe a prototype of an achromatic phase knife
coronagraph (Abe et al. \cite{abe2001a}) and its performance obtained in a laboratory experiment.
Second, we outline the installation and operating conditions of this coronagraph at the focus of
a $50$~cm refractor at the Observatoire de Nice (no central obstruction). We describe the
image-selection technique and present preliminary on-sky results obtained on the
binary star HD~80\,081 (*~38~Lyn). Possible ways to use such an instrument at Dome~C (Antarctica)
as a testbed instrument are also presented.

        \section{The achromatic hybrid phase mask coronagraph}
        \label{sec:AchroHybPhaseMaskCoro}

Following an earlier development, we decided to merge the two concepts of the achromatic phase knife
coronagraph (Abe et al. \cite{abe2001a}) and the four-quadrant phase mask coronagraph
(Rouan et al. \cite{rouan2000}), in order to simplify the design of the coronagraphic mask.

    \subsection{Phase mask design}
    \label{ssec:PhaseMaskDesign}

\begin{figure}
    \centering
    \begin{minipage}[c]{0.8\columnwidth}
        \centering\includegraphics[width=\columnwidth,clip=true]%
        {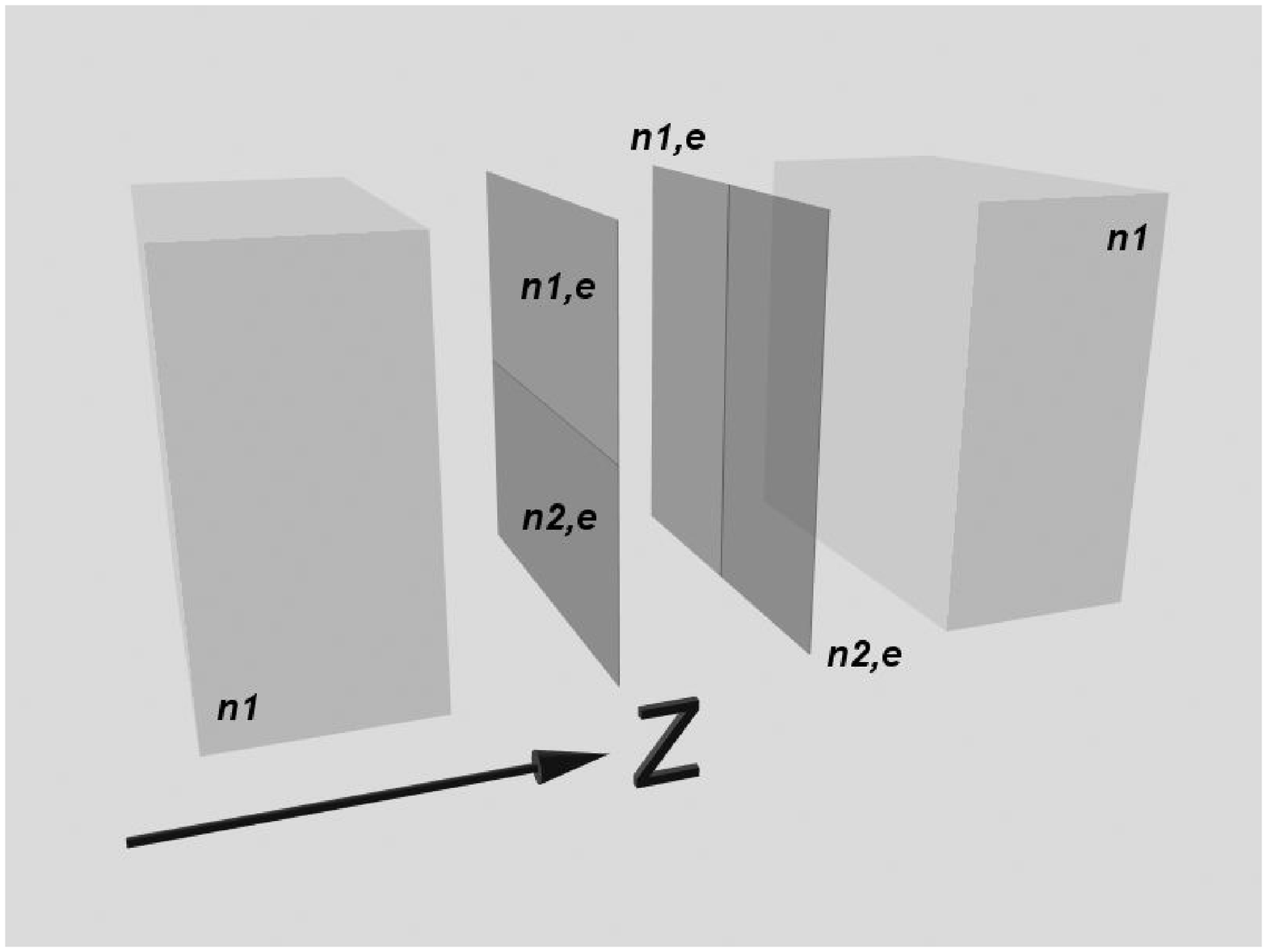}\\
        (a)
    \end{minipage}\\
    \begin{minipage}[c]{0.8\columnwidth}
        \centering\includegraphics[width=\columnwidth,clip=true]%
        {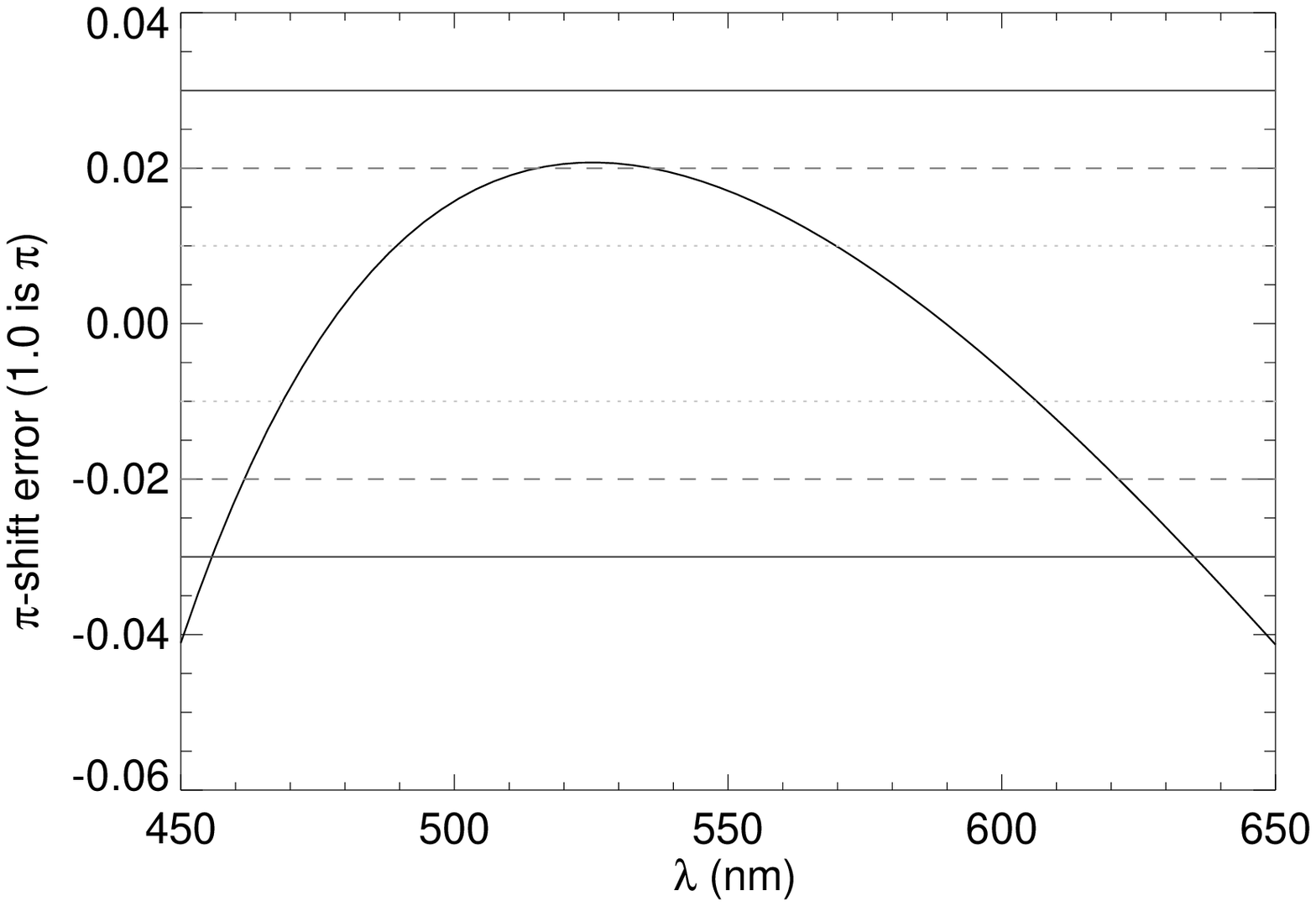}\\
        (b)
    \end{minipage}
   \caption[example]{
   \label{fig:APKC_specs}
   \emph{(a)}: The optical assembly of the
   APKC component: two $99$~$\mu$m phase knives are stacked between two
   $6$~mm thick glass plates. The light propagates along the $Z$ axis.
   \emph{(b)}: The relative error on the $\pi$ phase-shift over the considered
   spectral bandwidth.}
\end{figure}

The phase mask is an assembly of phase plates of equal thickness but different optical refraction indices.
Instead of using four separate plates, we used a combination of two crossed pairs, each one called a phase
knife (Abe et al. \cite{abe2001a}). The combination of indices ($n_1(\lambda)$ and $n_2(\lambda)$) and
glass thickness ($e_1=e_2=99$~$\mu$m) are optimized to obtain a theoretical nulling effect of $\sim10^4$
over a spectral band ranging from $\lambda=450$~nm to $620$~nm. The two orthogonal pairs of phase knives
are held together between two identical glass blocks (see Fig.~\ref{fig:APKC_specs}-a). This optical
design solution was chosen to circumvent the difficult operation of shaping and assembling four
independent quadrants.

Consider a light beam propagating through one phase knife. The
light passing through the left part of it will be phase-delayed
by the quantity:
\begin{equation}
  \varphi_1(\lambda) = \frac {2\pi e_1 n_1(\lambda)} {\lambda},
\end{equation}
and the second half of the wave will be phase-delayed by:
\begin{equation}
  \varphi_2(\lambda) = \frac {2\pi e_2 n_2(\lambda)} {\lambda}.
\end{equation}
For the sake of manufacturing and assembling simplicity, we impose
$e_1=e_2=e$. Therefore, the relative
phase-shift as a function of wavelength is given by:
\begin{equation}
  \Delta \varphi_{1,2}(\lambda) = \frac {2\pi e (n_1(\lambda) -
  n_2(\lambda))} {\lambda}.
\end{equation}
Figure~\ref{fig:APKC_specs}-b~shows the expected relative phase-shift error
$(\Delta \varphi_{1,2}-\pi)/\pi$ as a function of $\lambda$, computed for the chosen glass
combination: BaF4 and BaF52. The expected theoretical nulling effect of $\sim10^4$
requires that the differential glass thickness $\delta e$ be at most
$2$~nm ($\approx\:\lambda/300$).

    \subsection{Phase-mask fabrication}
    \label{ssec:PhaseMaskFab}

The two glass plates were first polished and molecularly bound to one of the polished glass blocks.
The glass plates were thick enough to prevent them from breaking while being manipulated.
Once bounded, the two plates were thinned to the specified thickness and then polished. As mentioned
previously, the differential thickness between the two plates should be less than a few nanometers,
which is not guaranteed by the present manufacturer. A similar assembly technique was used for the
second half of the coronagraphic mask.

The interface (edge contact) between the two glass plates is not well-defined and could be subject
to humidity infiltration. We thus chose to perform a V-shaped bevel at the plates interface
and to use an optical contact oil to seal it off once the component is assembled. The two separate
blocks were then assembled together using molecular bound. Finally, a UV-glue was used to coat
the component, holding the two parts together, ensuring a solid assembly.

        \section{Laboratory tests}
        \label{sec:LabTests}

    \subsection{Laboratory tests of the phase-mask}
    \label{ssec:LabTestsComp}
    
\begin{figure}
\centering
    \begin{minipage}[c]{\columnwidth}
        \centering\includegraphics[width=\columnwidth,clip=true]%
        {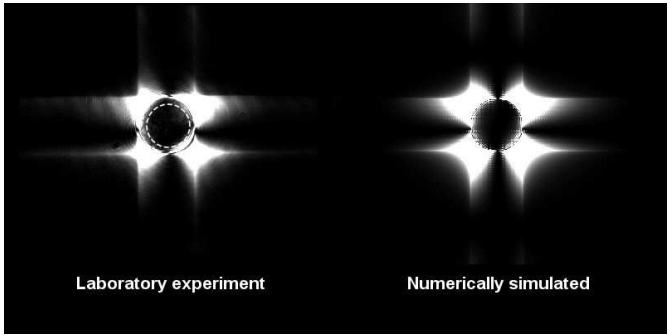}\\
    \end{minipage}
    \caption{
    \label{fig:APKC_pupil}
    Coronagraphic pupil images.
    \emph{Left}: laboratory experiment. The dashed circle represents the
    Lyot stop that is used for on-sky observations
    (see Section~\ref{ssec:OptoSetup}).
    \emph{Right}: numerically simulated. We introduced suspected
    phase-shift defects to simulate the pupil intensity distribution
    (thresholded images to exhibit the intensity distribution structures).}
\end{figure}

Although the component assembly was largely simplified using the
method described above, laboratory tests show that the molecular bound
between different glass indices may not be the optimal solution.
The left side of Figure~\ref{fig:APKC_pupil} shows the coronagraphic pupil image
obtained with the prototype. We identified a strong error on the
predicted $\pi$ phase-shift for one phase knife, indicating that
the molecular bound probably did not hold (\emph{e.g.} due to temperature
fluctuations). This effect might also be due to the polishing of
the two glass plates, which was not precise enough (thickness
difference). This phase-shift error can be reproduced by a numerical
simulation (right side of Figure~\ref{fig:APKC_pupil}) where we introduced an error of
$\sim10\,\%$ on the phase shift in one direction.

\begin{figure}
    \centering
    \begin{minipage}[c]{0.8\columnwidth}
        \centering\includegraphics[width=\textwidth,clip=true]%
        {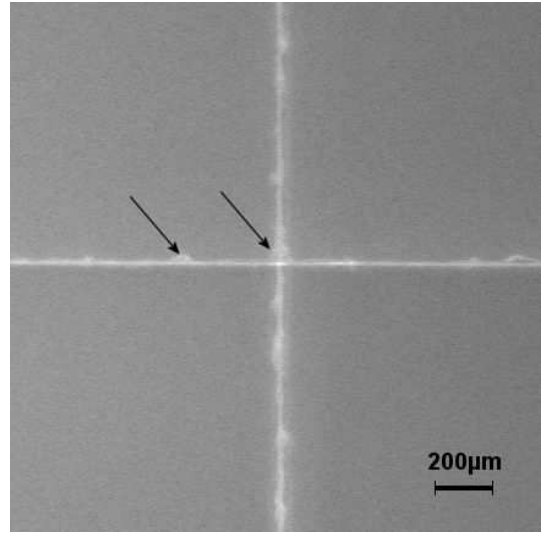}\\
    \end{minipage}
    \caption{
    \label{fig:cross_defects}
    Image of the coronagraphic mask illuminated by a diffuse white
    light source. Black arrows point toward defects (bubbles)
    produced by an optical oil used to avoid humidity infiltration.
    At the cross intersection, a faint defect is affecting the
    coronagraphic performance by modifying the image intensity
    distribution (see Fig.~\ref{fig:APKC_lab_perfs}-bottom right).}
\end{figure}

Another shortcoming of the coronagraph is shown in Figure~\ref{fig:cross_defects}. It turns out
that the optical sealing oil has created some bubbles that affect the regularity of the interface of
the phase-knives edge. It produces unwanted effects, especially at the center (see arrows in
Figure~\ref{fig:cross_defects}). The typical size of these irregularities is on the order of a few
tens of microns and is comparable to the size of the diffraction pattern.
This is probably the main reason the raw performance of the coronagraph is not as good as predicted
by the theory. The effect of these undesirable structures after assembling remains unclear,
although one can reasonably speculate that their diffraction effect in the coronagraphic pupil
is negligible compared to the expected turbulence degradation for on-sky observations.

        \subsection{Laboratory rejection measurements}
        \label{ssec:LabMeas}

\begin{figure}
    \centering
    \begin{minipage}[c]{0.55\columnwidth}
        \centering\includegraphics[width=\columnwidth,clip=true]%
        {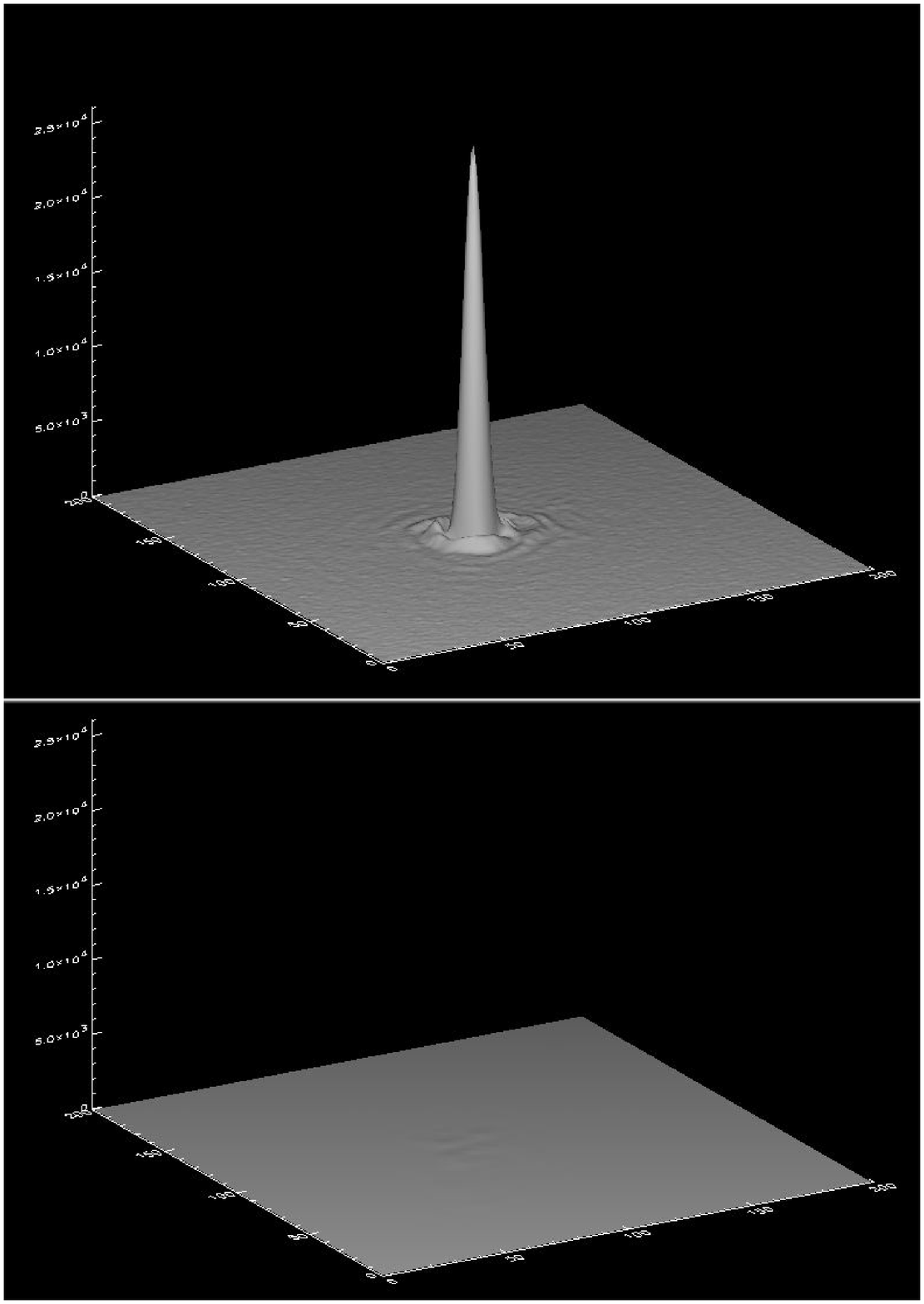}\\
    \end{minipage}
    \begin{minipage}[c]{0.40\columnwidth}
        \centering\includegraphics[width=\columnwidth,clip=true]%
        {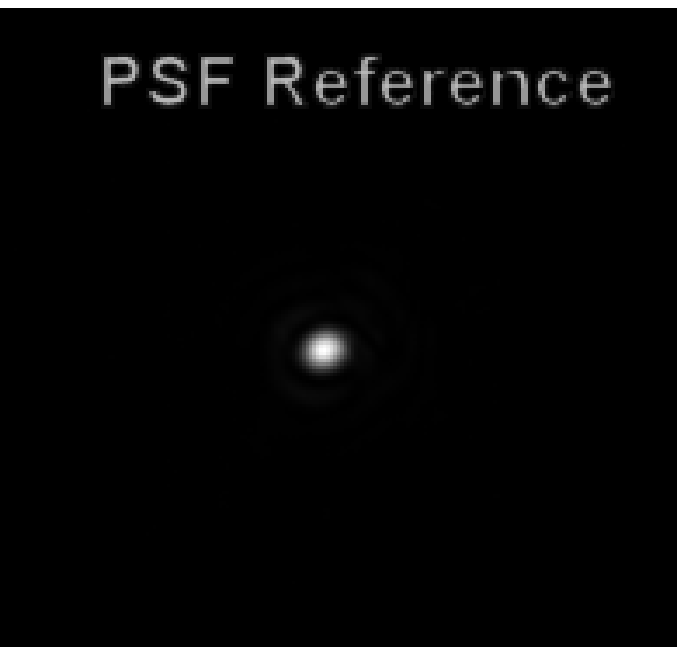}\\
        \centering\includegraphics[width=\columnwidth,clip=true]%
        {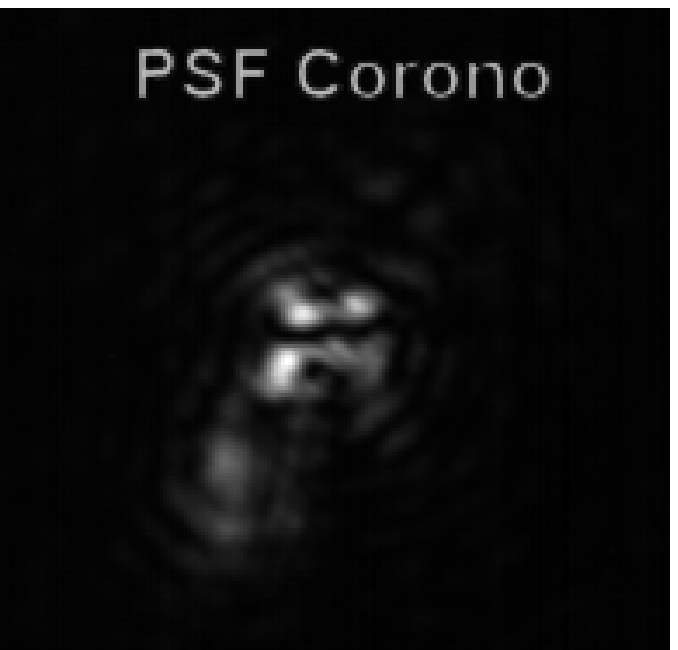}
    \end{minipage}
   \caption[example]{
   \label{fig:APKC_lab_perfs}
   \emph{Top-left}: the 3D representation of the non coronagraphic PSF with
   a linear intensity scale.
   \emph{Bottom-left}: the 3D representation of the coronagraphic PSF,
   normalized to the non-coronagraphic image peak intensity.
   \emph{Top- and bottom-right}: the corresponding 2D images, but with two
   different photometric scales, so that the coronagraphic image is visible.
   The peak-to-peak intensity ratio is $\sim1000$.}
\end{figure}

The coronagraphic bench was first qualified in laboratory prior to on-sky tests, under the same optical
conditions (defined $F/D=30$ ratio). The tests were made using an HeNe laser source ($632.8$~nm).
Due to the defect near the center mentioned in Section~\ref{sec:LabTests}, the nulling effect is less than
theoretically expected (a few tens of thousand extinction according to numerical models). The rejection
measuring procedure involves calibrated optical densities (Abe et al. \cite{abe2003}) to measure the
flux for the source on- and off-axis. This gave a peak-to-peak intensity ratio of
$\sim1000$ (Figure~\ref{fig:APKC_lab_perfs}) for the nulling performance.

        \section{Prior to on-sky tests}
        \label{sec:PriorOnSkyOnSky}

For on-sky observations, the coronagraph has been installed at the focus of the $50$~cm refractor of
the Observatoire de Nice, with $F=7.5$~m ($F/D=15$). However, an additional $25$~cm aperture diaphragm
turned out to be necessary, since we did not have any real-time turbulence correction system such as
an adaptive optics system or even a tip-tilt mirror.

The refractor objective lens is an achromatic doublet optimized for the $560$~nm wavelength. In our
case, the advantage of having no central obstruction had a counterpart: a residual chromatism strong
enough to hamper the image focus on the coronagraphic mask. Its effect on the coronagraphic rejection
was then evaluated numerically.

    \subsection{Numerical model for the refractor's residual chromatism effect}
    \label{ssec:SimuPerfs}

\begin{figure}
    \centering
    \begin{minipage}[c]{\columnwidth}
        \centering\includegraphics[width=\textwidth,clip=true]%
        {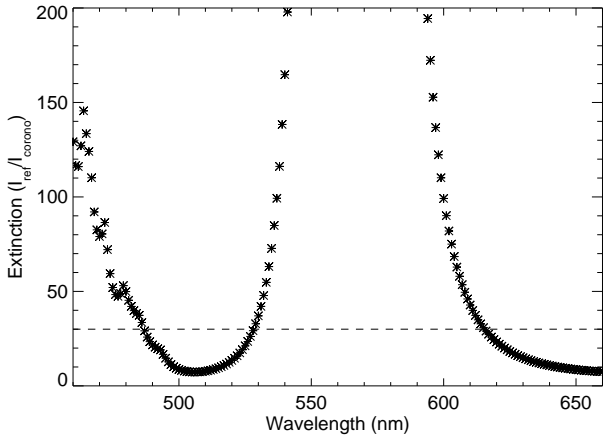}\\
    \end{minipage}
   \caption{
   \label{fig:effet_defocus_lunette}
Curve showing the wavelength-dependent extinction ($\max(I_{\mathrm{reference}})/\max(I_{\mathrm{coronagraph}})$)
(stars) due to the refractor focus chromatism and the extinction limit (dashed line) for the whole spectrum
(polychromatic coronagraphic image), which only reaches $\sim30$. The vertical scale has been truncated (because the
pure performance would be limited by the component itself to approximately 1000). A spectral filter would clearly
increase the extinction limit potential between 550~nm and 590~nm.}
\end{figure}

The graphs in Figure~\ref{fig:effet_defocus_lunette} are derived from a numerical simulation made with
the main characteristics of the refractor. It expresses the extinction performance at each wavelength.
The cumulated extinction is also given, while integrating the images over the whole spectral bandwidth.
Although the overall broadband extinction ratio performance is far from optimal (up to only
$\sim30$), we carried out a complete on-sky test under these conditions, essentially to match the
predicted observing parameters against real observing conditions.

        \subsection{Optical set-up for on-sky tests}
        \label{ssec:OptoSetup}

\begin{figure}
\centering
    \begin{minipage}[c]{\columnwidth}
        \centering\includegraphics[width=\textwidth,clip=true]%
        {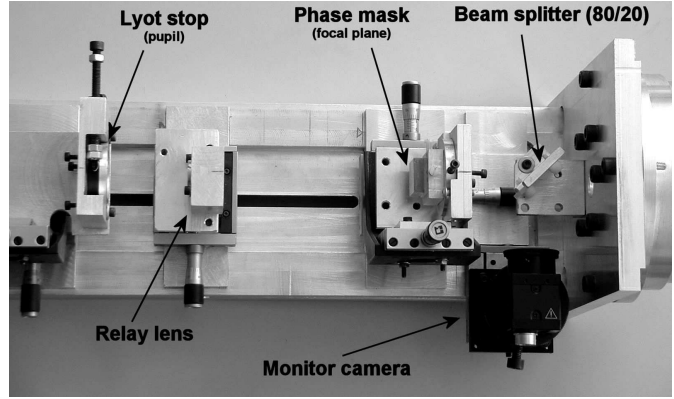}\\
    \end{minipage}
    \caption{
    \label{fig:Bench}
    Coronagraphic optical bench (see text for details). The refractor's bonnette
    is out of frame on the right side, and the science camera is out of frame
    on the left side. The light propagates from right to left.}
\end{figure}

Figure~\ref{fig:Bench} shows the coronagraphic optical bench
designed to be fitted to the bonnette of the $50$~cm refractor.
A beam-splitter reflects $20\,\%$ of the converging beam coming
from the refractor onto a monitoring camera (a SONY XC-EI50 CCD
camera with $768{\times}494$ $8.4{\times} 9.8$~$\mu$m pixels)
The remaining $80\,\%$ are transmitted through the coronagraph
to the science camera (a SONY XC-HR8500CE CCD camera with
$768{\times} 574$ $8.3$~$\mu$m square pixels, out of frame in
Figure~\ref{fig:Bench}). Both cameras deliver standard CCIR video
signals. An astigmatism aberration is introduced
by this beam-splitter. However, the phase mask is not sensitive
to it (Abe \cite{abe2002}) so that the coronagraphic performance
should not be degraded. The phase mask component is placed
at the focus of the refractor. A relay lens re-images both the
primary focal plane (onto the science camera's detector with a
magnification of $1$) and the pupil plane. At this relayed pupil
plane, an iris diaphragm (Lyot stop) with a diameter $20\,\%$ smaller
than the geometric image of the entrance pupil is placed to
eliminate the light diffracted out of the geometrical pupil (see
Fig.~\ref{fig:APKC_pupil}).

    \subsection{Acquisition software}
    \label{ssec:AcqSoft}

The acquisition software was
written in C++ with a graphical user interface (GUI). It drives a dual channel acquisition board
that accepts two standard CCIR signals with independent grabbers. The software handles the
simultaneous acquisitions from both cameras in real time with a multi-threaded architecture.
Several indicators can be optionally toggled in order to facilitate the fine tuning of the
experiment. There are two available modes:

\begin{itemize}
\item{}{A continuous recording of all the images in RAM memory: in
this mode, images from both cameras are recorded and saved in
memory, up to a capacity of $1.5$~GB. This mode is useful for dark
and flat recordings, or to assess the atmospheric turbulence
conditions (image motion for example).}
\item{}{An automatic image-selection mode: the user sets a threshold that has to be first
calibrated (see below). In this mode, memory is saved because only ``best'' images are kept,
so that the exposure length can be greatly optimized. The selection criteria are based on
flux considerations during the acquisition, although the post-acquisition image selection
was different (which will also be implemented in future experiments).}
\end{itemize}

In both modes, the millisecond timing of all acquired images are
recorded in separate text files in order to keep track of image-arrival
timing and avoid synchronization errors during image analysis.

Optionally, the software can send external correction commands to
the refractor guiding system, in order to keep the star centered
on the coronagraph axis. The frequency of correction mainly
compensates for slow drifts ($\sim1$~Hz).

        \section{Observing procedure}
        \label{sec:Calib}

Prior to the observations, the optical components were aligned to the bench's mechanical
axis in laboratory, so only minor fine tuning is necessary once it is plugged in to the
refractor. The focal mask and the Lyot stop longitudinal positions were tuned before starting
observations. The most effective phase mask position is found while monitoring the pupil image.
The resulting position is coherent with the focus position estimated using the Foucault method,
corrected from the focus shift introduced by the $6$~mm glass plate of the phase
mask (Figure~\ref{fig:APKC_specs}).

    \subsection{The image-selection process}
    \label{ssec:ImgSelect}

In this section, we propose an algorithm for image selection to reduce
the speckle noise in the coronagraphic image, and thus to improve
the signal to noise ratio for a companion detection.

\subsubsection{The speckle noise}
We call ``speckle noise'' the random speckle pattern due to distorted and/or tip-tilted wave-fronts.
These random structures become a noise source when the exposure time is not long enough to smoothe
them in a uniform halo. Unfortunately, the wavefront distortions induced either by atmospheric
turbulence and/or by instrumental aberrations do not generally smoothe out, since
they may have long time-scale variations (Marois et al. \cite{marois2003b}). These residual
speckles are the major limitations to the identification of the science target signal, especially
in the point-like source detection context (companion detection) since
speckles have the same size as the diffraction pattern. The
removal of such structures can be achieved by differential
techniques, but even in this case, residual differential defects
might still exist. The latter is discussed in several papers
(Racine et al. \cite{racine1999}; Marois et al. \cite{marois2003a};
Marois et al. \cite{marois2004}). Since the speckled halo is
partly correlated to the non
aberrated diffraction pattern (Bloemhof \cite{bloemhof2004};
Aime et al. \cite{aime2004}), its contribution can be diminished
by the use of a coronagraph.

In this context, image selection coupled to coronagraphy can be a
useful method of increasing the detection limit since the speckle
pattern will have a lower intensity due to the coronagraphic
attenuation, which is kept under a fixed level (because of
selection constraints) and therefore contributes to a lower
speckle variance at a given distance from the optical axis (see
Racine et al. \cite{racine1999}, Eq.~11b).

\subsubsection{The image-selection algorithm}
\label{sssec:SelectAlgo}
The image-selection process involves only the monitoring camera images, instead of relying
on flux considerations from the science camera (coronagraphic) image. It works as follows.
From each monitoring camera image, we compute the distance from the photometric
center\footnote{Standard centroid determination algorithms can be applied with a high
level of confidence, especially when the stellar companion is faint enough to be invisible
on individual snapshots.} of the object to a pre-computed operation point (OP).
When this distance exceeds a carefully chosen threshold, the corresponding coronagraphic
image from the science camera is rejected. The remaining images are co-added to produce the
``long-exposure'' coronagraphic image.

The OP has to be assessed accurately before each observation run by recording $5000$ frames
from both the monitoring and science cameras. First, a ``long-exposure'' image is obtained
by co-adding all the $5000$ science-camera frames. In the resulting image, the cross pattern
of the phase mask is clearly visible, and the optical center of the coronagraphic image can
be marked unambiguously.

Since the precise image selection procedure is not performed during the observation itself,
the OP on the monitoring camera is set according to its correlation with the images from
the science-camera having the highest extinction ratio.
For the post-data analysis, a refined determination of this correlation is performed.
To further improve the robustness of the proposed method, an internal (stable) calibration source
should be used to accurately set the OP and even to allow precise image selection in real-time. In
the present work, the determination of the OP position could only be set on a pixel scale basis
(with an image selection radius of 0.5 pix at best).

        \section{Application to observation data}
        \label{sec:ApplData}

On the nights of observations, the atmospheric conditions could not be considered as optimal.
Despite rapid seeing variations, presumably due to fast high-altitude winds, a coherent peak
was almost always present in every short exposure. Diffusion from high-altitude clouds was
also perceptible, with a noticeable effect (halo) on the stacked images (see below).

    \subsection{Observation of HD~80\,081 (binary star)}
    \label{ssec:ObsLynx}

\begin{figure}
    \centering
    \begin{minipage}[c]{0.45\columnwidth}
        \centering\includegraphics[width=\textwidth,clip=true]%
        {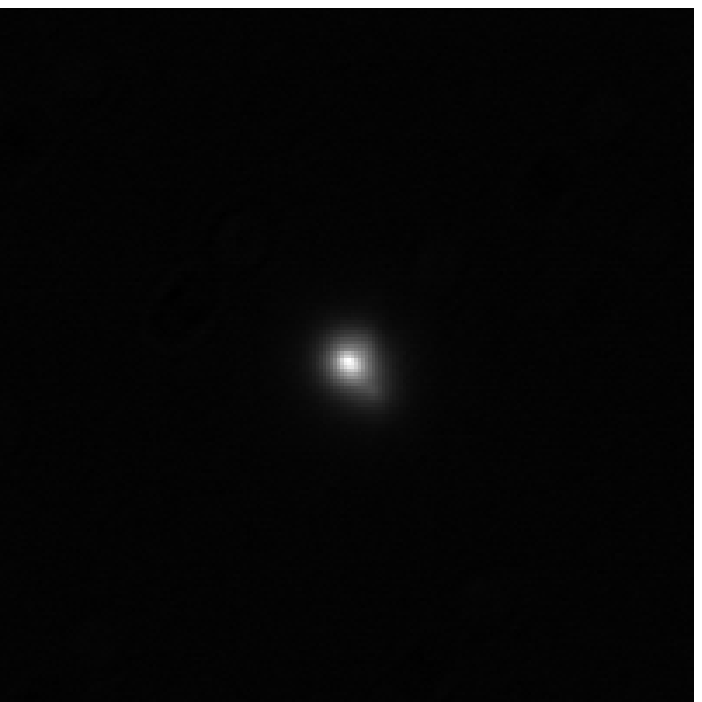}\\
        (a)
    \end{minipage}
    \begin{minipage}[c]{0.45\columnwidth}
        \centering\includegraphics[width=\textwidth,clip=true]%
        {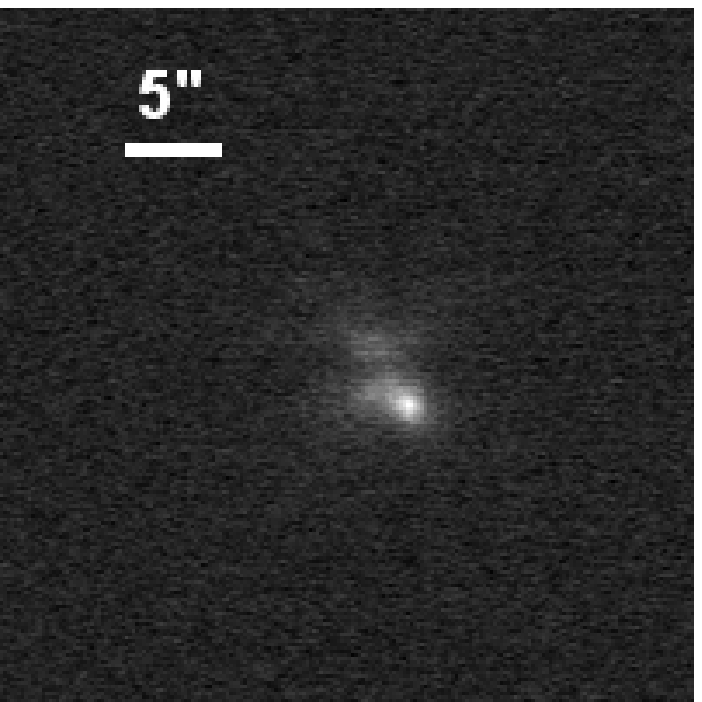}\\
        (b)
    \end{minipage}
   \caption{
   \label{fig:obs_data}
   On-sky data of HD~80\,081.
   \emph{(a)}: without the coronagraph (monitoring camera),
   \emph{(b)}: with the coronagraph (science camera). This image was obtained
   by co-adding $126$~frames selected among the $5000$ recorded frames, using
   the selection algorithm described in Sect.~\ref{ssec:DataCubeAnalysis}.
   The companion appears clearly in the coronagraphic image.}
\end{figure}

The data under consideration in this article were obtained on March 3, 2004. The target we chose
for this test of methodology was HD~80\,081, a binary star with a primary magnitude $V_1=3.9$,
secondary magnitude $V_2=6.6$ ($\Delta V=2.7$), and separation $2.7\arcsec$. The orientation
on the sky of the secondary component was known, and the coronagraph was oriented so that it
didn't focus on a ``blind area'' of the phase mask (i.e. $\sim45\degr$ relative to the mask
axis). Indeed, investigating the unknown morphology and complexity of a target is hampered by
the  ``blind areas'' of the four quadrant phase mask, adding a real difficulty to data analysis.
Note that the secondary component was not visible in single science-camera snapshots, due to
the read-out noise (S/N less than unity), and was hardly visible with the monitoring camera. As
already mentioned, no real-time guiding correction was active on the refractor, so that the
image selection post-processing became mandatory. For comparison, Figure~\ref{fig:obs_data}
shows the monitor camera exposure and the corresponding coronagraphic image. Image generation
and analysis are detailed in the following paragraphs.

    \subsection{Data cube analysis}
    \label{ssec:DataCubeAnalysis}

\begin{figure*}
    \centering
    \begin{minipage}[c]{0.32\textwidth}
        \centering\includegraphics[width=\textwidth,clip=true]%
        {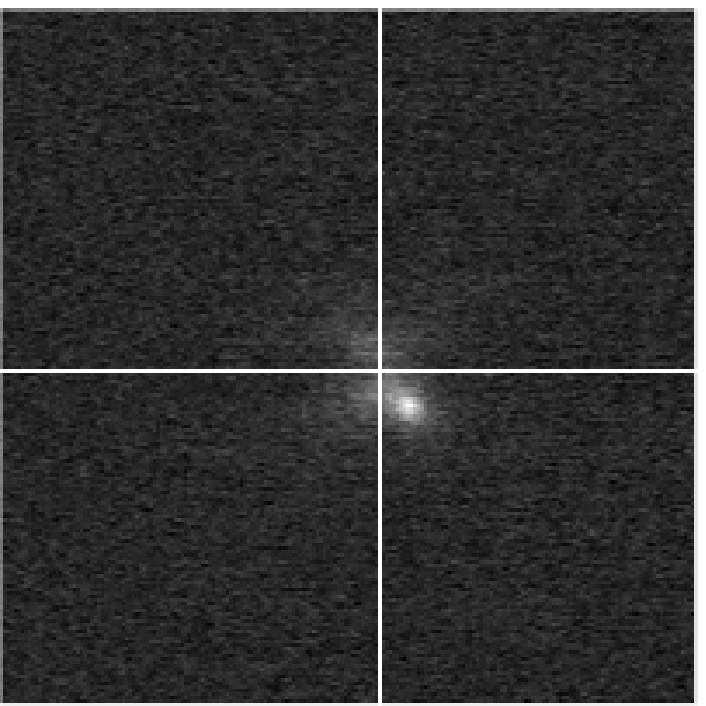}\\
        (a)
    \end{minipage}
    \begin{minipage}[c]{0.32\textwidth}
        \centering\includegraphics[width=\textwidth,clip=true]%
        {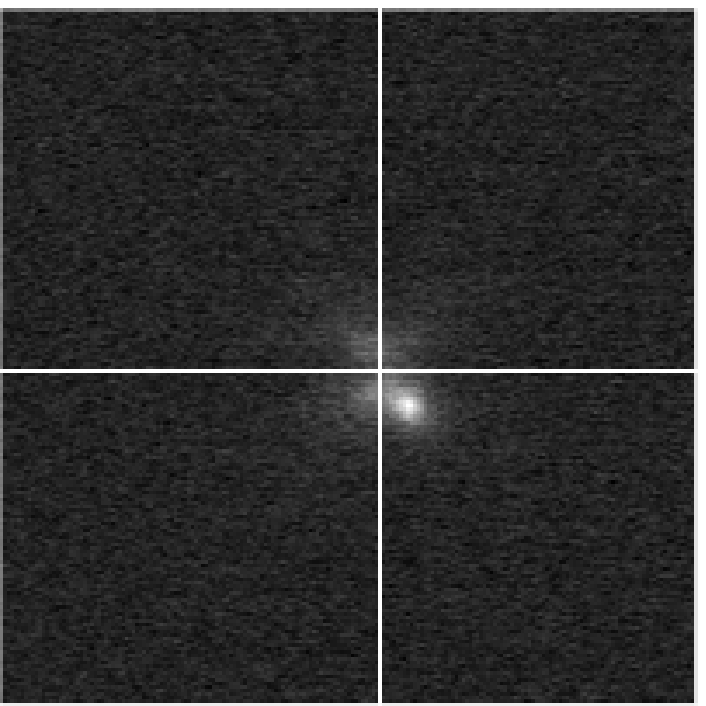}\\
        (b)
    \end{minipage}
    \begin{minipage}[c]{0.32\textwidth}
        \centering\includegraphics[width=\textwidth,clip=true]%
        {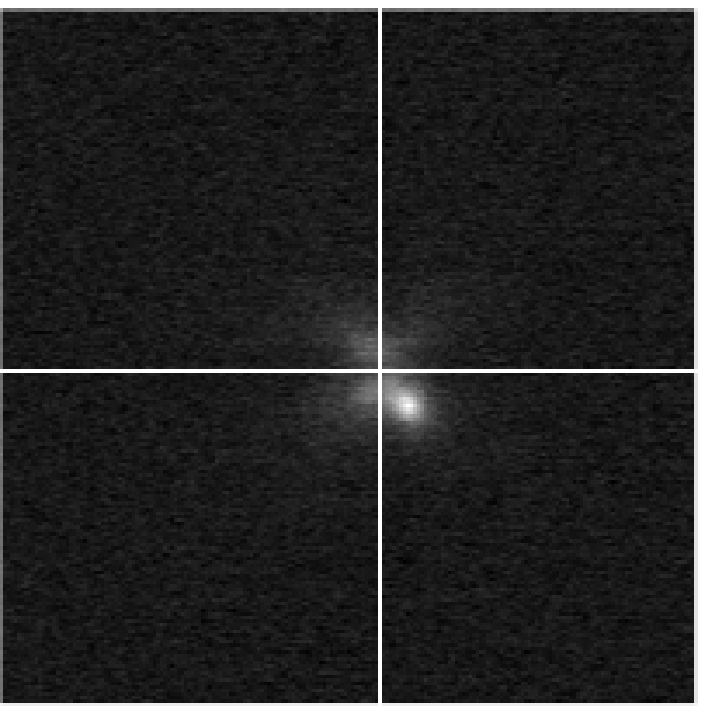}\\
        (c)
    \end{minipage}
    \begin{minipage}[c]{0.32\textwidth}
        \centering\includegraphics[width=\textwidth,clip=true]%
        {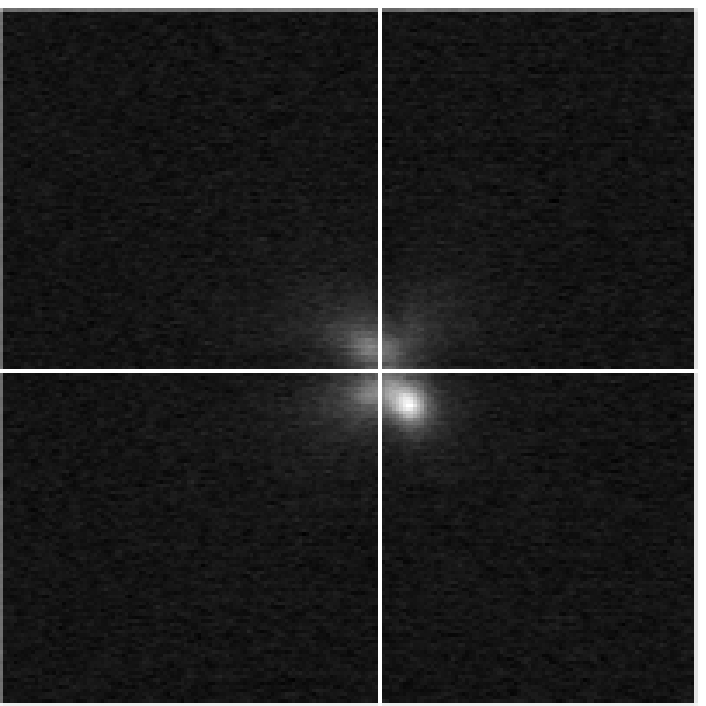}\\
        (d)
    \end{minipage}
    \begin{minipage}[c]{0.32\textwidth}
        \centering\includegraphics[width=\textwidth,clip=true]%
        {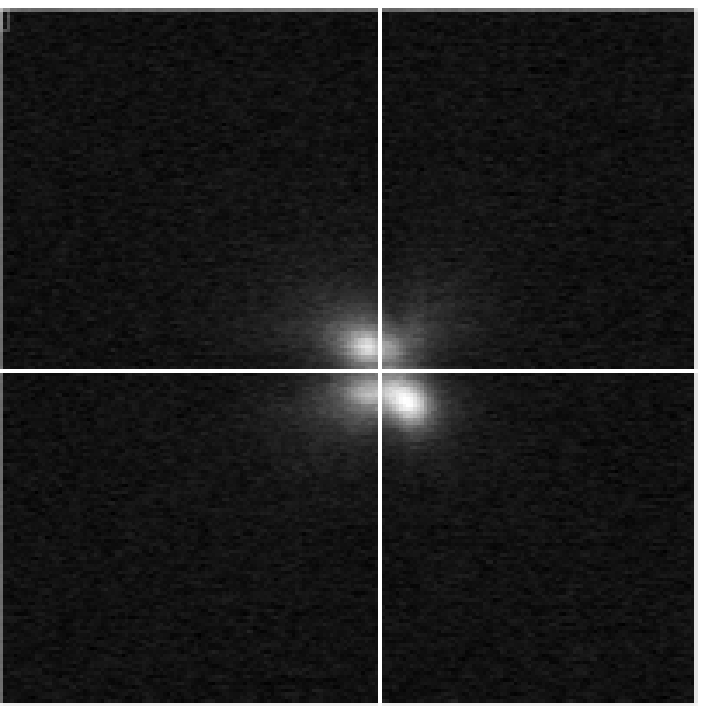}\\
        (e)
    \end{minipage}
    \begin{minipage}[c]{0.32\textwidth}
        \centering\includegraphics[width=\textwidth,clip=true]%
        {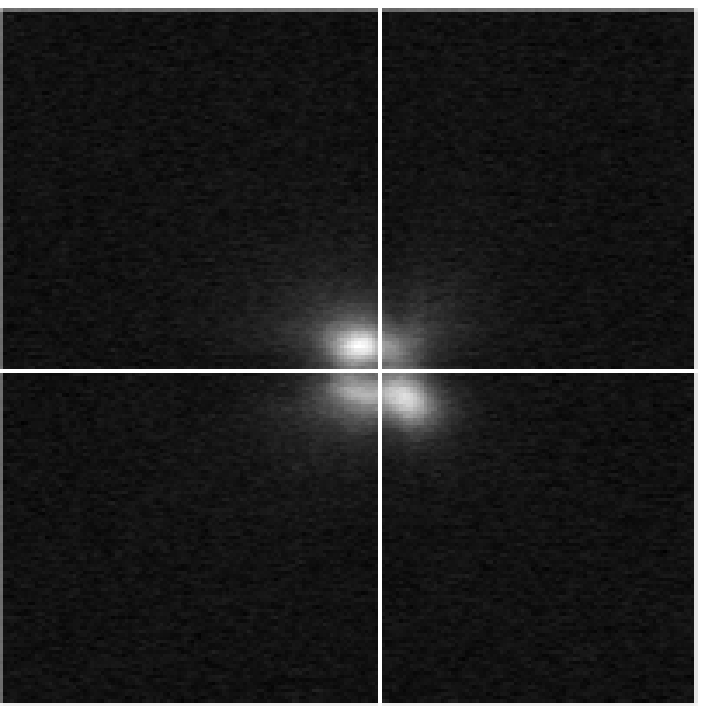}\\
        (f)
    \end{minipage}
    \caption{
    \label{fig:corono_images_select}
    Coronagraphic images generated with increasing values of the selection
    distance threshold $\eta$ (in pixels and in units of $\lambda/D$)
    and thus with an increasing number of retained frames. The residual speckles
    become brighter, thus reducing the companion detection S/N.
    \emph{(a)}: $\eta=0.5$ px $=0.25\,\lambda/D$\, ($\sim80$ frames);\,\,
    \emph{(b)}: $\eta=0.7$ px $=0.35\,\lambda/D$\, ($\sim130$ frames);\,\,
    \emph{(c)}: $\eta=1.0$ px $=0.5\,\lambda/D$\, ($\sim250$ frames);\,\,
    \emph{(d)}: $\eta=2.0$ px $=1.0\,\lambda/D$\, ($\sim900$ frames);\,\,
    \emph{(e)}: $\eta=4.0$ px $=2\,\lambda/D$\, ($\sim2750$ frames);\,\,
    \emph{(f)}: $\eta=8.0$ px $=4\,\lambda/D$\, ($\sim4500$ frames).}
\end{figure*}

We performed the image selection procedure described in Sect.~\ref{ssec:ImgSelect}, with increasing
values of the distance threshold and thus increasing numbers of retained frames.
Figure~\ref{fig:corono_images_select} displays the resulting coronagraphic images.

In order to check the efficiency of the proposed selection method, we compared the peak intensity
of the HD~80\,081 secondary component with the peak intensity of the brightest residual speckle
(Figure~\ref{fig:brightness_compare_speckle}-b). This ratio of intensity is plotted as a function
of the selection threshold in Figure~\ref{fig:brightness_compare_speckle}-a. This illustrates the gain
provided by the selection method which overcomes two effects:

\begin{figure}
    \centering
    \begin{minipage}[c]{0.5\columnwidth}
        \centering\includegraphics[width=\textwidth,clip=true]%
        {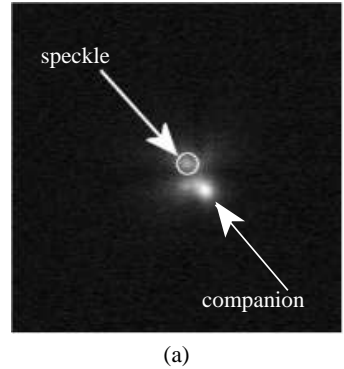}\\
        (a)
    \end{minipage}
    \begin{minipage}[c]{\columnwidth}
        \centering\includegraphics[width=\textwidth,clip=true]%
        {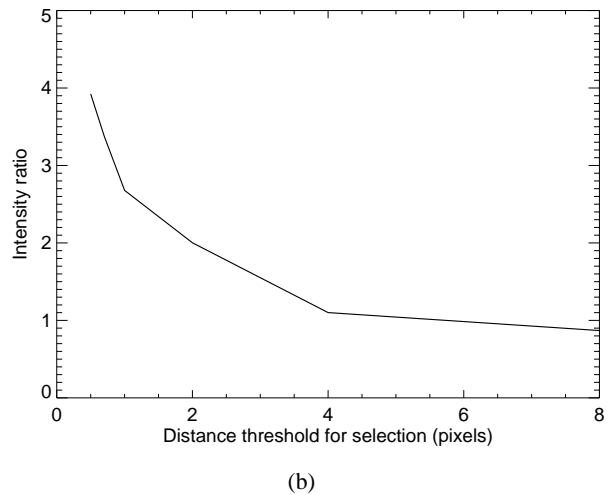}\\
        (b)
    \end{minipage}
   \caption{
   \label{fig:brightness_compare_speckle}
   \emph{(a)}: The final coronagraphic image showing the brightest
   residual speckle of HD~80\,081 (emphasized by a white circle surrounding it)
   and the clearly visible companion of HD~80\,081.
   \emph{(b)}: The intensity ratio between HD~80\,081's companion and
   the brightest residual speckle after coronagraphy,
   as a function of the distance threshold in the image selection process
   (see Sect.~\ref{sssec:SelectAlgo}).}
\end{figure}

\begin{itemize}
\item the apparition of bright speckles due to unconstrained tip-tilt jitters that move
the primary star's image out of the coronagraphic rejection zone;
\item the flux dilution of the target peak intensity that tends to lower
the difference between the target peak and the residual bright speckles mentioned above.%
\end{itemize}

These effects are clearly visible in the Figure~\ref{fig:corono_images_select} images, where
in the last long exposure (almost all frames are integrated), the companion of HD~80\,081
cannot be disentangled from the coronagraphic image halo with the additional confusion that
another bright speckle, symmetric to the HD~80\,081 companion position has appeared.

The peak-to-peak intensity ratio between the coronagraphic and direct image was measured
to approximately $30$, which is consistent with the numerical estimate mentioned in
Sect.~\ref{ssec:SimuPerfs} (see Figure~\ref{fig:effet_defocus_lunette}).

\subsection{Data consistency checks}
\begin{figure}
    \centering
    \begin{minipage}[c]{\columnwidth}
        \centering\includegraphics[width=\textwidth,clip=true]%
        {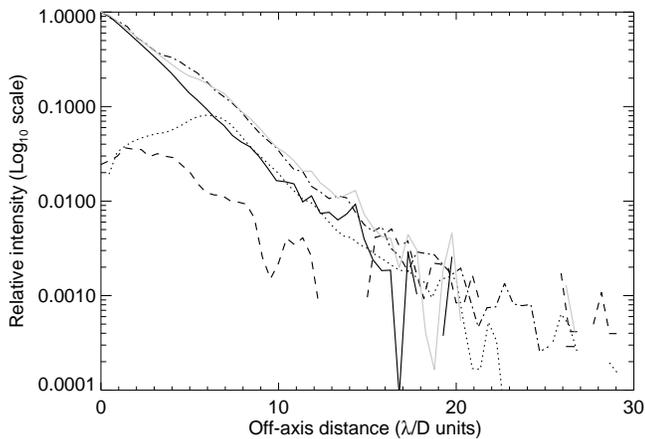}\\
    \end{minipage}
   \caption{
   \label{fig:profils_images_compare}
   Comparison of image profiles derived from exposure (b) of
   Figure~\ref{fig:corono_images_select}. See text for details.}
\end{figure}

To complete the data reduction, we also checked the result consistency with the long exposure image.
Figure~\ref{fig:profils_images_compare} shows several radial profiles for exposure (b) of
Figure~\ref{fig:corono_images_select} with $126$ selected frames and a selection radius of $0.7$~pixels:
\begin{itemize}
\item
the radial median profile of the non-coronagraphic monitor image (solid curve);
\item
the average profile of the the monitor image along the direction of the off-axis
companion\footnote{we averaged the radial profiles within a small sector containing
the secondary component of HD~80\,081l.} (dash-dotted curve);
\item
the same with the coronagraphic image (dotted curve);
\item
the radial median profile of the coronagraphic image (dashed curve);
\item
the sum of solid and dotted curves (light grey curve),
which is consistent with the dash-dotted curve.
\end{itemize}
This comparison allows us to confidently conclude that our coronagraphic data are consistent and that the
bright speckle is the companion of HD~80\,081. These results clearly show that the central star PSF halo
has been attenuated and that the star is not simply masked by the coronagraphic mask. If there were no
coronagraphic effect, then the intensity distribution at the companion location would be similar to the
dash-dotted curve. The attenuation factor at the companion location ($\sim 6 \lambda/D$) is $\approx 8$,
while it reaches $\sim 10$ closer to the optical axis ($\sim 4 \lambda/D$), where the theoretical
transmission of the four quadrant phase mask is $\approx 90\%$.

        \section{Conclusion and future prospects}
        \label{sec:Conclusion}

We have presented the first on-sky results of a visible achromatic hybrid phase knife coronagraph.
The extinction performances, limited by the refractor's residual chromatism and thus modest,
were found to agree with the numerically predicted value ($\sim30$).

We demonstrated on real data the efficiency of the image selection algorithm we propose in order to
increase the signal to noise ratio and ease the detection of faint companions.
The selection method provides an efficient way to extract the scientific information by preventing
bright speckles from appearing and confusing to the image interpretation.

This encouraging experiment is now envisaged for a test instrument at Dome~C (Antarctica). The instrument
is expected to benefit from diffraction-limited images while the current image selection system
is being implemented. Furthermore, we plan to combine this optical system with a high-speed,
high-efficiency photon-counting camera that is under construction at the ``Laboratoire Universitaire
d'Astrophysique de Nice'' (LUAN). We will therefore benefit from quasi-absent readout noise and
good temporal sampling of the atmospheric turbulence. The host instrument will be a 14-inch C14
telescope with a modified entrance pupil (Lloyd et al. \cite{lloyd2003}) in order to keep a
maximal coronagraphic performance despite the central obscuration (with a lower transmission).

In addition to this gain in transparency, seeing conditions and chromatism reduction (due
to the use of a reflector), this new device at Dome~C will allow long integration periods
during the polar night. This mission is also a test experiment
after the success of smaller optical instruments already tested at
Dome~C for monitoring the seeing (Aristidi et al. \cite{aristidi2003}).

\begin{acknowledgements}
The authors would like to thank G{\'e}rard~GREISS (SEOP, France) for
manufacturing the achromatic phase mask, Alex~ROBINI from LUAN who
manufactured most of the mechanical parts, and OCA and ASHRA for
supporting the experiment. Lyu ABE is grateful to the CNES for its
support during his postdoctoral fellowship.
This prototyping experiment was supported by a special grant from
the scientific council of the Observatoire de la C{\^o}te d'Azur.
\end{acknowledgements}

\end{document}